# Designed Three-Dimensional Freestanding Single-Crystal Carbon Architectures

Ji-Hoon Park[1]**, Dae-Hyun Cho[2]**, Youngkwon Moon[1], Ha-Chul Shin[1], Sung-Joon Ahn[1], Sang Kyu Kwak[3], Hyeon-Jin Shin[4], Changgu Lee[2,5]*, and Joung Real Ahn[1,5]*

[1]Department of Physics, Sungkyunkwan University, Suwon 440-746, Republic of Korea

[2]Department of Mechanical Engineering, Sungkyunkwan University, Suwon 440-746, Republic of Korea

[3]School of Energy and Chemical Engineering, Ulsan National Institute of Science and Technology (UNIST), Ulsan 689-798, Republic of Korea

[4]Graphene Center, Samsung Advanced Institute of Technology, Yongin 446-712, Republic of Korea

[5]SKKU Advanced Institute of Nanotechnology, Sungkyunkwan University, Suwon 440-746, Republic of Korea

* To whom correspondence should be addressed: E-mail: peterlee@skku.edu(CL), jrahn@skku.edu(JRA),

**These authors contributed equally to this work.

**Abstract**

Single-crystal carbon nanomaterials have led to great advances in nanotechnology. The first single-crystal carbon nanomaterial, fullerene, was fabricated in a zero-dimensional form. One-dimensional carbon nanotubes and two-dimensional graphene have since followed and continue to provide further impetus to this field. In this study, we fabricated designed three-dimensional (3D) single-crystal carbon architectures by using silicon carbide templates. For this method, a designed 3D SiC structure was transformed into a 3D freestanding single-crystal carbon structure that retained the original SiC structure by performing a simple single-step thermal process. The SiC structure inside the 3D carbon structure is self-etched, which results in a 3D freestanding carbon structure. The 3D carbon structure is a single crystal with the same hexagonal close-packed structure as graphene. The size of the carbon structures can be controlled from the nanoscale to the microscale, and arrays of these structures can be scaled up to the wafer scale. The 3D freestanding carbon structures were found to be mechanically stable even after repeated loading. The relationship between the reversible mechanical deformation of a carbon structure and its electrical conductance was also investigated. Our method of fabricating designed 3D freestanding single-crystal graphene architectures opens up prospects in the field of single-crystal carbon nanomaterials, and paves the way for the development of 3D single-crystal carbon devices.

Single-crystal carbon nanomaterials, such as zero-dimensional (0D) fullerene, one-dimensional (1D) carbon nanotubes, and two-dimensional (2D) graphene, have extraordinary electrical, optical, and mechanical properties, and so have triggered the rapid development of nanotechnology.[1-5] For example, the high carrier mobility of graphene has contributed to the development of high speed electronic devices, such as radio-frequency transistors,[6-8] and its high transmittance and flexibility mean that it has found applications in flexible and transparent devices.[9,10] Fullerene, carbon nanotubes, and graphene intrinsically prefer 0D, 1D, and 2D geometries, respectively, so their application has been limited to the replacement of planar parts of 0D, 1D, and 2D devices, such as electrodes and channel materials. Recently, three-dimensional (3D) carbon networks have been fabricated in attempts to overcome these geometrical limitations.[11,12] 3D carbon networks were first grown with chemical vapor deposition (CVD) using nickel foams as templates, and graphene-based cellular monoliths have been fabricated with freeze-casting processes.[13,14] These 3D carbon networks have high specific surface areas and retain the electrical conductance, elasticity, and flexibility of planar carbon structures. It has been suggested that the high specific surface areas of 3D carbon networks mean that they can be used as electrode materials in lithium ion batteries, supercapacitors, and channel materials in gas sensors.[15-17]

The fabrication of 3D carbon networks has resulted in significant advances in graphene research. However, the porous structures of 3D carbon networks are polycrystalline. Furthermore, such porous structures cannot be used in 3D electronic devices, as discussed below. Recently, 2D electronic devices have been replaced by 3D electronic devices with multiple stacked and/or vertical geometries, and with freestanding and/or supported geometries, such as 3D hybrid complementary metal-oxide-semiconductors (CMOSs) or 3D nanoelectromechanical systems (NEMSs).[18,19] 3D electronic structures have been developed in order to enhance integration density and to facilitate the lateral downsizing of electronic devices. Such architectures have also resulted in better performance, higher connectivity, reduced interconnect delays, lower power consumption, better space utilization, and flexible heterogeneous integration. To use graphene in such electronic devices, designed 3D graphene architectures need to be realized. For this reason, the porous structures of 3D carbon networks are not appropriate for 3D electronic devices. Therefore, a different approach to the fabrication of designed 3D carbon architectures is required. In particular, it remains challenging to realize 3D freestanding single-crystal carbon architecture. The ability to fabricate such architectures will lead to the development of 3D single-crystal carbon devices, and will open up prospects in the field of single-crystal carbon nanomaterials.

In our study, we demonstrated that 3D freestanding single-crystal carbon architecture can be designed and fabricated, and that the method used can be extended from nanoscale to microscale. A designed 3D single-crystal SiC wafer was used to grow designed 3D freestanding carbon architecture (Figure 1). For our method, when the SiC wafer is heated at a high temperature in an argon atmosphere,[20-22] the 3D SiC architecture is etched, leaving a 3D freestanding carbon architecture that resembles the original 3D SiC template. This was confirmed with scanning tunneling microscopy (SEM), Raman spectroscopy, and atomic force microscopy (AFM). Furthermore, the hollowness of the 3D graphene architecture was confirmed by examining the underlying SiC structures after the selective removal of the carbon. Interestingly, the 3D carbon architecture was found to be single-crystal, with the same hexagonal close-packed structure as graphene, using low energy electron diffraction (LEED) (Figure 2). These results demonstrate that designed 3D freestanding single-crystal carbon architectures can be grown by using a simple single-step process without further transfer and/or etching processes. Furthermore, the designed 3D freestanding single-crystal carbon architecture persisted after mechanical loading, as confirmed with AFM and molecular dynamics (MD) simulations. This mechanical stability also means that the passage of electrical current through the 3D carbon architectures is reversible with respect to their deformation, as demonstrated with conductive AFM measurements. Finally, 3D carbon architecture was fabricated in an isolated form by transferring it onto a SiO$_2$ wafer.

**RESULTS AND DISCUSSION**

Figure 1 shows representative false color field-emission SEM images of the designed 3D freestanding carbon architectures. The designed 3D freestanding carbon architecture has the same hexagonal close-packed structure as graphene. Thus, we will call it designed 3D freestanding graphene architecture hereafter. The 3D SiC architecture shaped like cylindrical pillars was fabricated with electron-beam lithography and reactive ion etching (Figures 1a and c); each pillar-shaped structure has a radius and height of 240 nm and 800 nm, respectively. After resistive heating at 1750°C under an argon pressure of 180 Torr, the 3D SiC architecture was transformed into 3D freestanding graphene architecture (Figures 1b, d and e). As shown in Figures 1d and e, the 3D graphene architecture is transparent to the electron beam, which suggests that it is hollow and freestanding. Furthermore, we were able to control the size of the designed 3D freestanding graphene architectures from the nanoscale to the microscale. The designed 3D freestanding graphene architecture was fabricated from templates obtained with various lithographical methods, namely electron lithography, photolithography, and focused ion beam. For example, four different truncated cone-shaped SiC architectures were fabricated with a focused ion beam; these SiC structures had radii of 0.1, 0.15, 0.25, and 0.5 µm (Figure 1f). These 3D SiC structures were transformed into freestanding graphene architectures that retained the original geometries of the templates, as shown in Fig. 1g. When the designed SiC templates were heated at temperatures that deviated from the optimal temperature for the growth of 3D freestanding graphene architectures, 3D freestanding graphene did not grow (Supplementary Information Figure S1). Up to 1600°C, graphene does not grow and the SiC architecture is not altered, as shown by the optical microscopy images (Supplementary Information Figures S1a and e). At 1650°C, the SiC architecture is etched without the growth of graphene (Supplementary Information Figures S1b and e). Graphene starts growing at 1700°C and this growth is optimized at 1750°C (Supplementary Information Figures S1c-e).

To determine the wafer-scale crystallinity of the 3D freestanding graphene architecture, LEED with an electron beam size of approximately 1 mm was used. An array of microscale 3D SiC structures shaped like inverted bowls was fabricated over an entire wafer with photolithography.[23] Figure 2a shows a false color SEM image of the 3D freestanding graphene architecture that resulted from the inverted bowl-shaped 3D SiC architecture. Figure 2b shows a representative LEED pattern of the 3D freestanding graphene architecture that was reproduced over the entire wafer.[24] Interestingly, this LEED pattern demonstrates that the 3D freestanding graphene architecture is single-crystalline; this single crystallinity is maintained at all positions on the wafer.

To further understand the growth mechanism, as well as the mechanical and electrical properties, of the 3D freestanding graphene architecture, microscale freestanding graphene architecture was examined. Figures 3a-d show a schematic of the growth process of the 3D freestanding graphene architecture during heating at 1750°C under an argon pressure of 180 Torr. Initially, graphene grows over the surface of the 3D SiC architecture (Figure 3a). Epitaxial graphene on a SiC substrate grows when Si atoms are sublimated at high temperatures and the remaining C atoms are bonded together to produce graphene.[20-22] For the 3D SiC architecture, Si atoms on the surface of the architecture are sublimated. Graphene is sequentially grown on the surface of the architecture so that the structure of the 3D graphene resembles that of the 3D SiC architecture. After growing the graphene, the SiC inside the 3D graphene architecture is etched, resulting in freestanding 3D graphene architecture (Figures 3b-d). The rate of Si sublimation is an important parameter, and it increases at a step structure in comparison to a terrace structure.[25,26] The 3D SiC structures are composed of microscopic step structures, while the plates at the bottom are made up of terraces. The high rate of Si sublimation at the 3D structures is because the step structures may temporarily trap the sublimated Si atoms between the graphene and the SiC. These trapped Si atoms may start the etching of the remaining C atoms and hinder the formation of graphene. When the etching process occurs, the etching will be accelerated because many defects are created and the rate of Si sublimation increases. This growth mechanism was confirmed by etching the graphene and examining the remaining SiC at each stage (Figures 3g-i). Inverted bowl-type SiC architecture was fabricated with photolithography (Figure 3e); the inverted bowls had a diameter of 5 μm and a height of 500 nm. Figure 3f shows a SEM image of the 3D graphene architecture that was grown from the 3D SiC architecture in Figure 3e. To observe the SiC structure underlying the graphene architecture, the graphene was selectively etched with oxygen plasma, which does not damage the SiC structures (Supplementary Information Figure S2). After etching the graphene architecture, the remaining SiC structures clearly show that they are gradually etched over time and are finally flattened (Figures 3g-i). The selective etching of graphene demonstrates that the 3D graphene architecture is freestanding.

The hollowness of the freestanding graphene architecture was further confirmed with Raman spectroscopy and AFM (Supplementary Information Figures S3 and S4). The Raman spectra were obtained from the top of the 3D architecture at various stages of the heating process (Supplementary Information Figure S3). The intensities of the SiC peaks with Raman shifts between 1000 and 2000 cm$^{-1}$ gradually decrease while the intensities of the typical Raman peaks of graphene, the G (1591.5 cm$^{-1}$) and 2D peaks (2710.5 cm$^{-1}$), gradually increase.[27,28] The G and 2D peaks originate from the breathing modes of $sp^2$ carbon atoms and two phonons with opposing momentum in the highest optical branch near the K point of the Brillouin zone (BZ) of graphene, respectively.[29] The changing intensities of the Raman peaks also suggest that the SiC structure inside the graphene architecture is etched. The ratio of the intensities of the 2D and G peaks, $I_{2D}/I_G$, is approximately 1 and is almost independent of the heating time. This intensity ratio suggests that the graphene architecture has the characteristics of bilayer graphene. The D peak of graphene is mostly caused by defects or disordered structures. In our case, it has a low intensity that suggests the 3D freestanding graphene has a low defect density.

The mechanical response of the 3D graphene architecture to indentation was also measured to confirm its hollowness (Supplementary Information Figure S4). Curves of mechanical load against indentation depth were determined at each stage of heating (Supplementary Information Figure S4). An AFM tip can indent soft graphene but not a solid SiC structure. The load-depth curves show that the mechanical load is suddenly enhanced when the AFM tip begins to touch the underlying SiC structure. The curve reaches points of high stiffness at depths of approximately 10, 90, and 350 nm for heating times of 40 seconds, 4 minutes, and 7 minutes, respectively. These changes in the critical mechanical load suggest that the SiC architecture underlying the 3D graphene architecture is gradually etched after the growth of graphene, resulting in 3D freestanding graphene architecture. This is consistent with the SEM images obtained after the selective etching of graphene. Therefore, several different experiments have confirmed that the 3D graphene architecture is freestanding.

The mechanical stability of the designed 3D graphene structures was further assessed with AFM. Figure 4 shows the mechanical responses produced by indenting the center of the 3D freestanding graphene structures with an AFM tip. Diamond-like carbon-coated AFM tips with lengths less than 15 nm were used to avoid geometric or chemical changes in the tips during indentation. Using the reference cantilever method, the spring constant of the cantilever was calibrated with respect to the spring constant of a pre-calibrated cantilever provided by BRUKER (CLFC-NOMB).[30] To assess the mechanical stability of the graphene structures, AFM topographic images were obtained using the tapping mode after the structures were indented. Figure 4a shows an AFM image and a line profile of an inverted bowl-shaped 3D freestanding graphene structure before indentation. Figures 4e-g show AFM images and line profiles that were acquired after multiple indentations of varying depth. These results suggest that after multiple indentations, the overall geometry of the 3D freestanding graphene structure is maintained. The AFM experiments demonstrate that the resilience of the graphene structures is sufficient to bear repetitive mechanical loads with almost complete recovery. The hysteresis loops shown originate from the difference between the loads at the onset of the abrupt decrease in the slope during loading and unloading. The hysteresis corresponds to approximately 46% of the energy absorbed during the indentation cycle. Typically, nonlinearity and sudden changes in the slope observed during loading and/or unloading processes can be attributed to the viscoelasticity or damping behavior of the materials[31,32] and the buckling of structures (Supporting information Figure S5). [33]

As described above, the designed 3D freestanding single-crystal graphene structures can endure repeated large compressive stresses with excellent strain recovery. Furthermore, this method of fabricating graphene structures has the advantage of controlling their size from the nanoscale to the microscale because the templates used can be successfully fabricated with either electron lithography or photolithography. Therefore, this approach to the engineering of 3D nanoscale or microscale graphene architectures enables the development of 3D materials. When the superior properties of graphene are combined with designed 3D architectures, a wide range of applications can be achieved, such as heat transfer enhancement,[34] the control of surface wettability,[35] the development of strain sensors,[36] and the fabrication of coatings that absorb mechanical energy.[37] As an example of these applications, we examined the dependence of the electrical conductivity of a graphene structure on the load exerted on it to explore its applicability as a pressure (or touch) sensor. A conductive chromium/platinum-coated AFM tip was used to measure the variation of electric conductance during the loading-unloading cycle. Figure 5 shows the variation of electrical conductance with applied load. The electrical conductance was found to be reversible during the loading-unloading cycle. This reversibility indicates that 3D freestanding graphene architectures can be used in pressure (or touch) sensors.

The 3D freestanding graphene architectures can be directly fabricated on top of a SiC wafer, as described above. To examine whether the physical properties of the 3D graphene architectures can be preserved after being transferred to other substrates, a $SiO_2$/Si wafer was used. Figures S6a and b in the Supplementary Information show SEM and optical images, respectively, of a 3D graphene architecture transferred onto a $SiO_2$/Si wafer. A thermal release tape was used to transfer the graphene.[38] Interestingly, the optical and SEM images suggest that the 3D geometry can be maintained during the transfer process and can be stabilized on other substrates. These images also indicate that the mechanical stability of the 3D freestanding graphene architectures is intrinsic. Furthermore, the D peaks in the Raman spectra show that no additional defects are produced by the transfer process (Supplementary Information Figure S6c). The relative intensity of the 2D and G peaks of the Raman spectra ($I_{2D}/I_G$) is inversely proportional to the number of layers of graphene present.[29] The 3D graphene had $I_{2D}/I_G$ ratios ranging from 4.0 to 0.5, but most of the 3D graphene had $I_{2D}/I_G$ ratios of approximately 1. As previously reported for graphene transferred to a $SiO_2$ film,[39] monolayer graphene has an $I_{2D}/I_G$ ratio greater than 2 and bilayer graphene has an $I_{2D}/I_G$ ratio of 1. Therefore, the number of graphene layers in the 3D structures ranges from monolayer to trilayer. Most of the 3D graphene had $I_{2D}/I_G$ ratios of approximately 1, so the majority of the 3D graphene structures were composed of bilayer graphene. The number of graphene layers can be roughly controlled by adjusting the heating temperature. When the 3D graphene was produced at the optimized temperature of 1750°C, the 3D graphene was bilayer. Monolayer and trilayer graphene was made when the heating temperatures were close to 1700°C and above 1750°C, respectively. Curves of mechanical load against indentation depth were also recorded (Supplementary Information Figures S6d-f). After multiple indentations, the line profiles were determined (Supplementary Information Figures S6g-i). Surprisingly, there was no noticeable degradation of the structures with respect to that of the as-grown architecture. The observed resilience also suggests that the mechanical stability of the designed 3D freestanding graphene architecture is intrinsic to the structures.

MD simulations were performed to investigate the reversible deformation processes of the designed 3D freestanding graphene architectures. We modeled a 3D freestanding graphene structure on the atomic scale to capture its loading-unloading behavior (Figure 6). The system consisted of a Si cone, a 3D freestanding graphene structure, and a 6H-SiC substrate, which consisted of 310, 3975, and 12288 atoms, respectively. Figures 6a-e show snapshots obtained at intervals of 15.7 ps. Since the total energy of the modeled system is conserved during the simulation, the complete transfer of momentum from the moving cone to the freestanding graphene is expected. Thus, the interaction potential energy of the freestanding graphene was measured instead of calculating the loading force directly. There are bonding and non-bonding energy terms in the potential energy: bonding valence and cross terms and the non-bonding van der Waals term. It was found that the majority of the potential energy belongs to valence energy terms (approximately 98.5%). These terms are a result of the direct interactions of bonded atoms ($sp^2$-hybridized carbons), such as bending, stretching, and torsion. Even on this small scale, our simulation replicated the experimental hysteresis by producing a clear discrepancy between the potential energy curves (Figure 6f). This supports the idea that the mechanical hysteresis is intrinsic to the 3D freestanding graphene structure, which is governed by direct bonding when under strain.

## CONCLUSIONS

Designed 3D freestanding single-crystal carbon architectures have been successfully fabricated. 3D carbon architectures were grown from 3D single-crystal SiC templates fabricated with various lithographic methods. The hollowness of the 3D carbon architecture was confirmed with various experiments. The 3D carbon architecture was also mechanically stable, as demonstrated with AFM indentation measurements and MD simulations. We expect that this method for the fabrication of designed 3D freestanding single-crystal carbon architecture will lead to the development of new types of 3D single-crystal carbon devices with diverse and novel functions.

## METHODS

**Preparation of 3D graphene architecture.** A 3D SiC structure was patterned on a n-type 6H-SiC(0001) wafer (Cree Inc) by dry etching via photolithography. The mask used a chromium/quartz glass template with a 6 μm dot pattern and 8 μm pitch. The open area of the SiC wafer after photolithography was etched out by reactive ion etching using $SF_6$/Ar mixed gases, where the etching rate was approximately 500 Å /min.[23] Then, the masking photoresist was stripped with acetone, and the patterned structure was confirmed by optical spectrometer. Then, graphene was epitaxially grown at 1750 °C under an argon pressure of 180 Torr using resistive heating.

**Characterization of 3D graphene architecture.** The formation of epitaxial graphene on SiC was monitored in situ using low-energy electron diffraction (LEED, Specs). Scanning electron microscopy (SEM) images of the 3D patterned SiC and transferred graphene were acquired using a field-emission SEM (JEOL JSM7500F, 15kV). The exact size of the 3D graphene transferred onto a $SiO_2$ substrate was measured using atomic force microscopy (AFM, SeikoSPA 400) in DFM mode. Raman spectra were measured to determine the number of layers and the quality of the transferred graphene using a micro-Raman spectroscopy (Renishaw, RM1000-Invia, 514 nm, $Ar^+$ ion laser).

**SUPPORTING INFORMATION AVAILABLE:** Figure S1-S6. This material is available free of charge *via* the Internet at http://pubs.acs.org

**ACKNOWLEDGEMENTS**

This study was supported by a National Research Foundation of Korea (NRF) grant (No. 2012R1A1A2041241) and the National Research Foundation (NRF) of Korean grant funded by the Korean Ministry of Science, ICT and Planning (No. 2012R1A3A2048816). CL was supported by the Center for Advanced Soft Electronics under the Global Frontier Research Program (20110031629) funded by the Korean government and a National Research Foundation of Korea (NRF) grant (No. 2009-0083540)

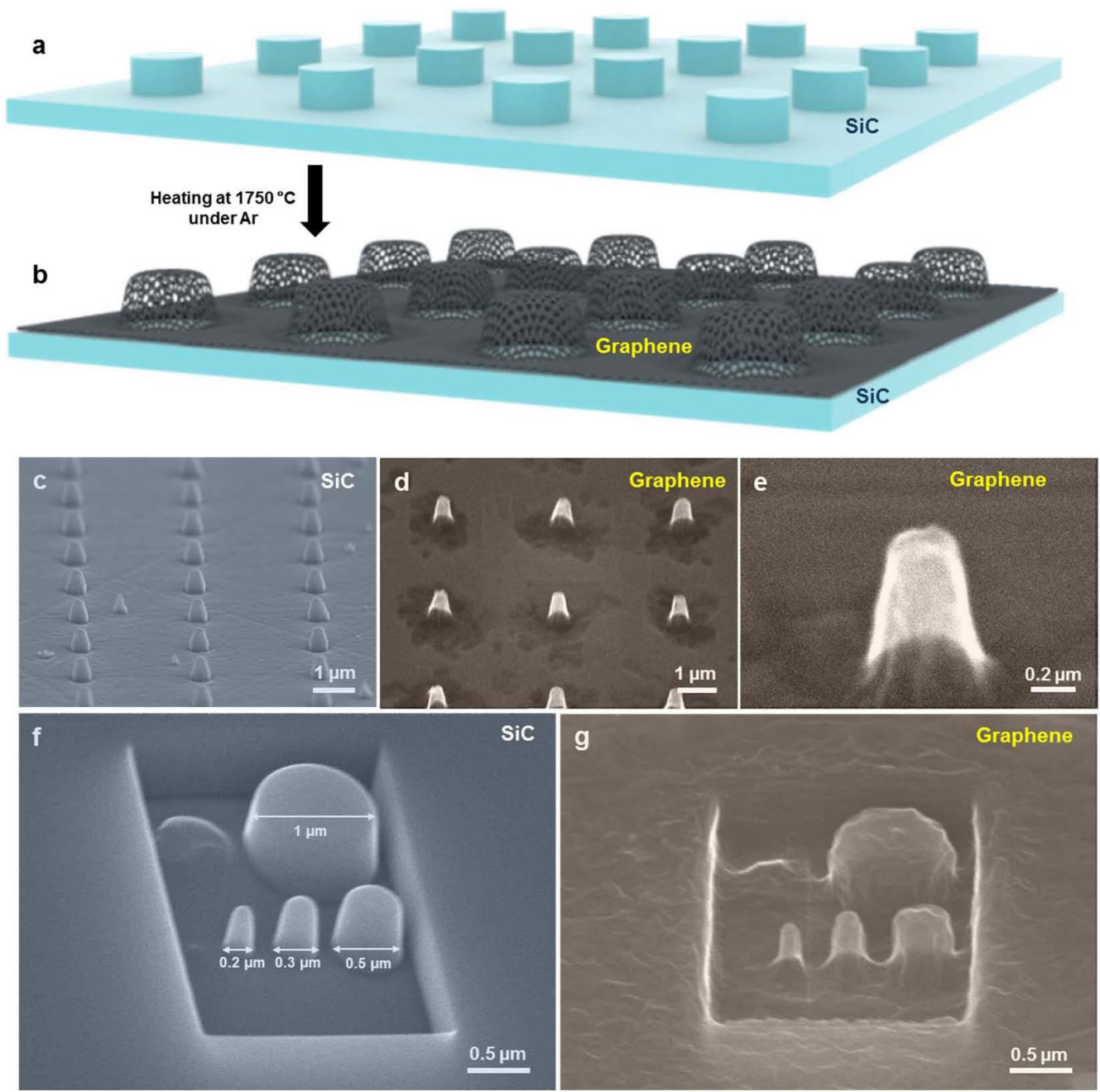

**Figure 1 | SEM images of 3D SiC and freestanding 3D graphene structures**. (**a-b**) A schematic of the fabrication of designed 3D freestanding graphene architecture using a designed 3D SiC architecture as a template. (**c**) A false color SEM image of pillar-shaped SiC structures fabricated with electron lithography. (**d**) A false color SEM image of the pillar-shaped 3D graphene structures resulting from **c**. (**e**) An

enlarged SEM image of **d**. (**f**) A false color SEM image of various SiC architectures fabricated with a focused ion beam. (**g**) A false color SEM image of the 3D graphene architectures obtained from the templates in **f**.

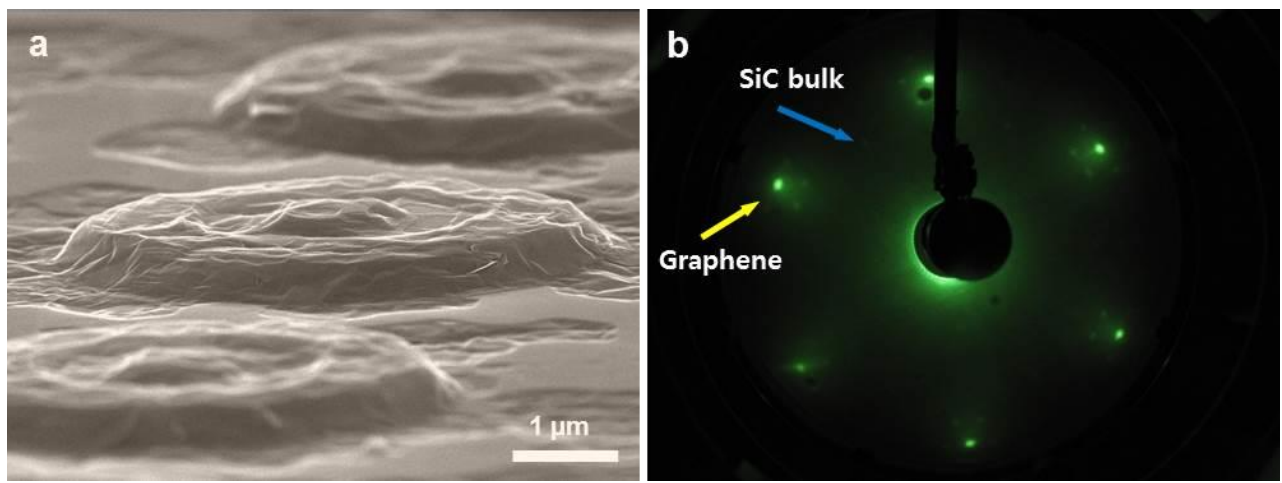

**Figure 2 | Crystallinity of a 3D freestanding graphene structure**. (**a**) A false color SEM image of 3D freestanding graphene structures shaped like an inverted bowl. (**b**) The LEED pattern produced by the graphene architecture.

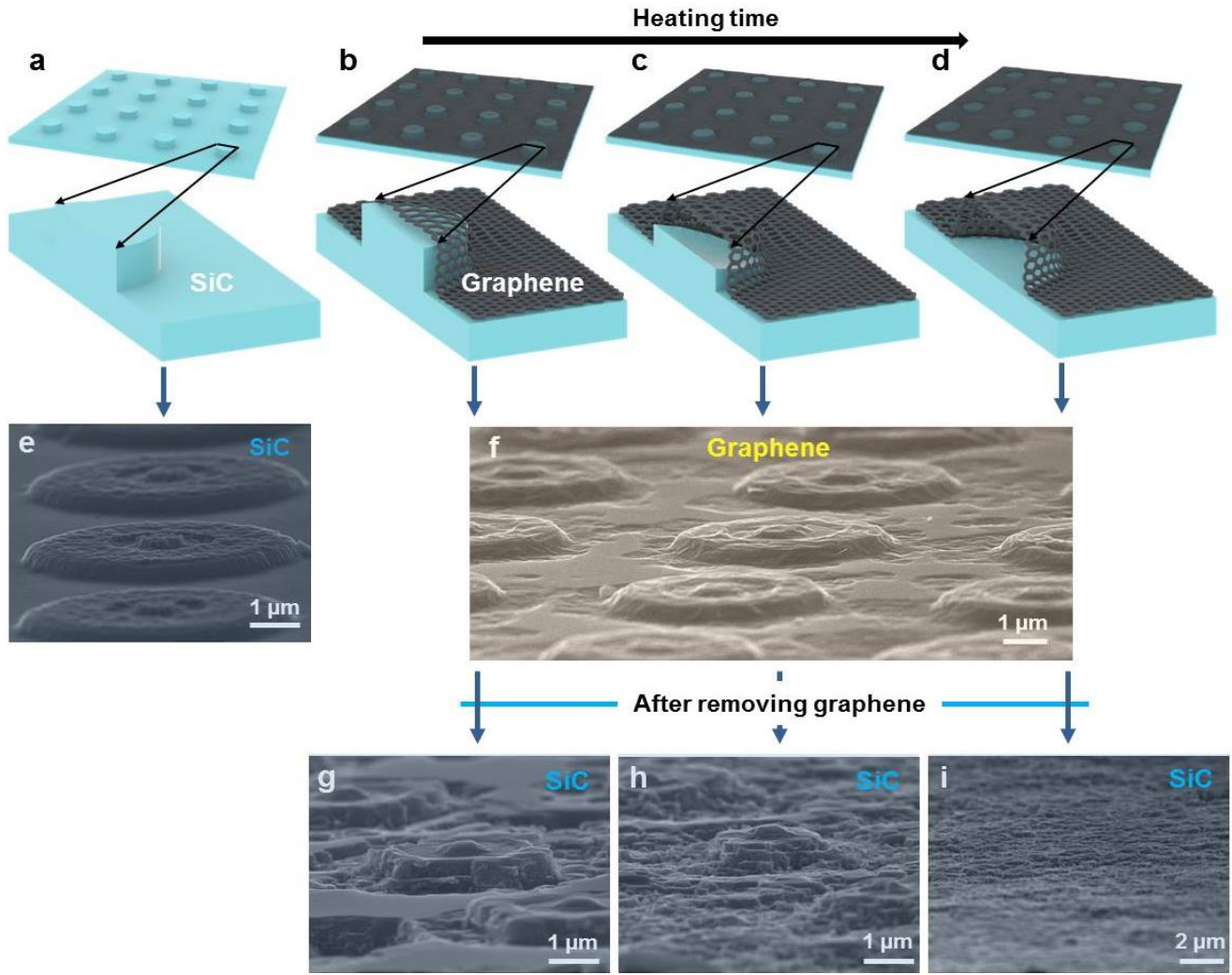

**Figure 3 | Growth process of 3D freestanding graphene architecture**. (**a-d**) Schematic of the growth mechanism of the 3D freestanding graphene architecture, where the blue and black regions represent SiC and graphene, respectively. (**e**) A SEM image of a designed SiC template. (**f**) A SEM image of the 3D graphene architecture grown on the SiC template. (**g-h**) SEM images of the SiC structures underlying the 3D graphene architecture, which were exposed by removing the 3D graphene architecture with oxygen plasma. The images in **g**, **h**, and **i** correspond to those in **b**, **c**, and **d**, respectively.

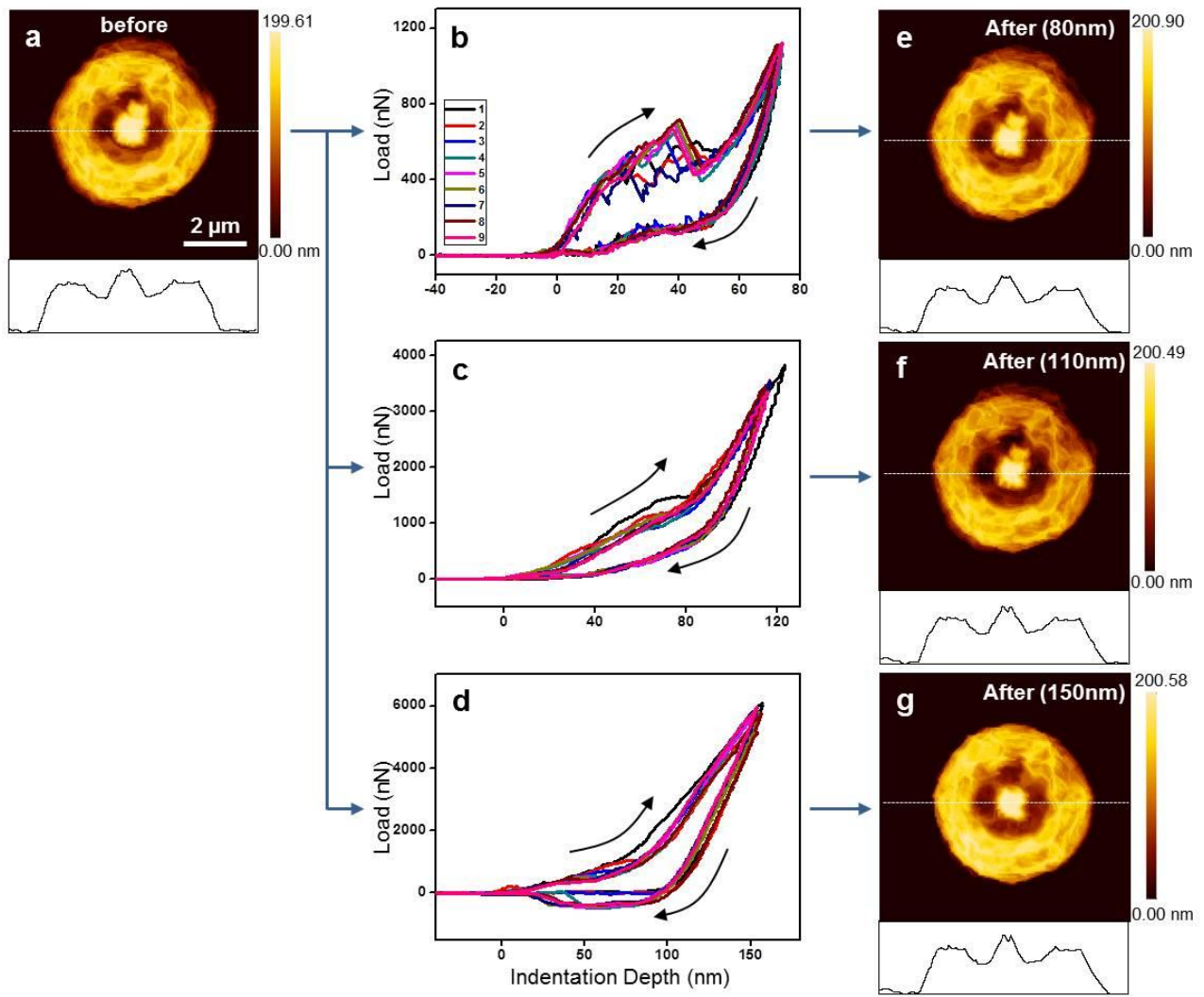

**Figure 4 | Mechanical properties of a 3D freestanding graphene structure**. (**a**) AFM topography and a line profile along the dotted line of an inverted bowl-shaped graphene structure before indentation. (**b-d**) Load-displacement curves of the first 9 cycles at indentation depths of 80 nm (**b**), 110 nm (**c**), and 150 nm (**d**). (**e-g**) AFM topographies and line profiles of the graphene structure after the indentations.

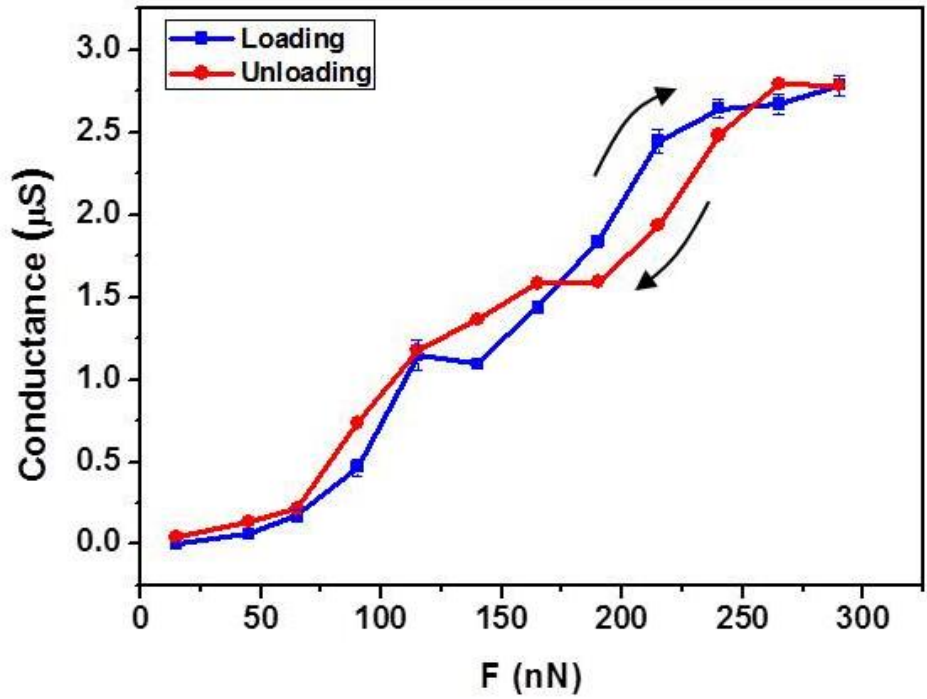

**Figure 5 | Electrical properties of 3D freestanding graphene**. The variation in the electrical conductance of the inverted bowl-shaped 3D freestanding graphene structure during a loading and unloading cycle.

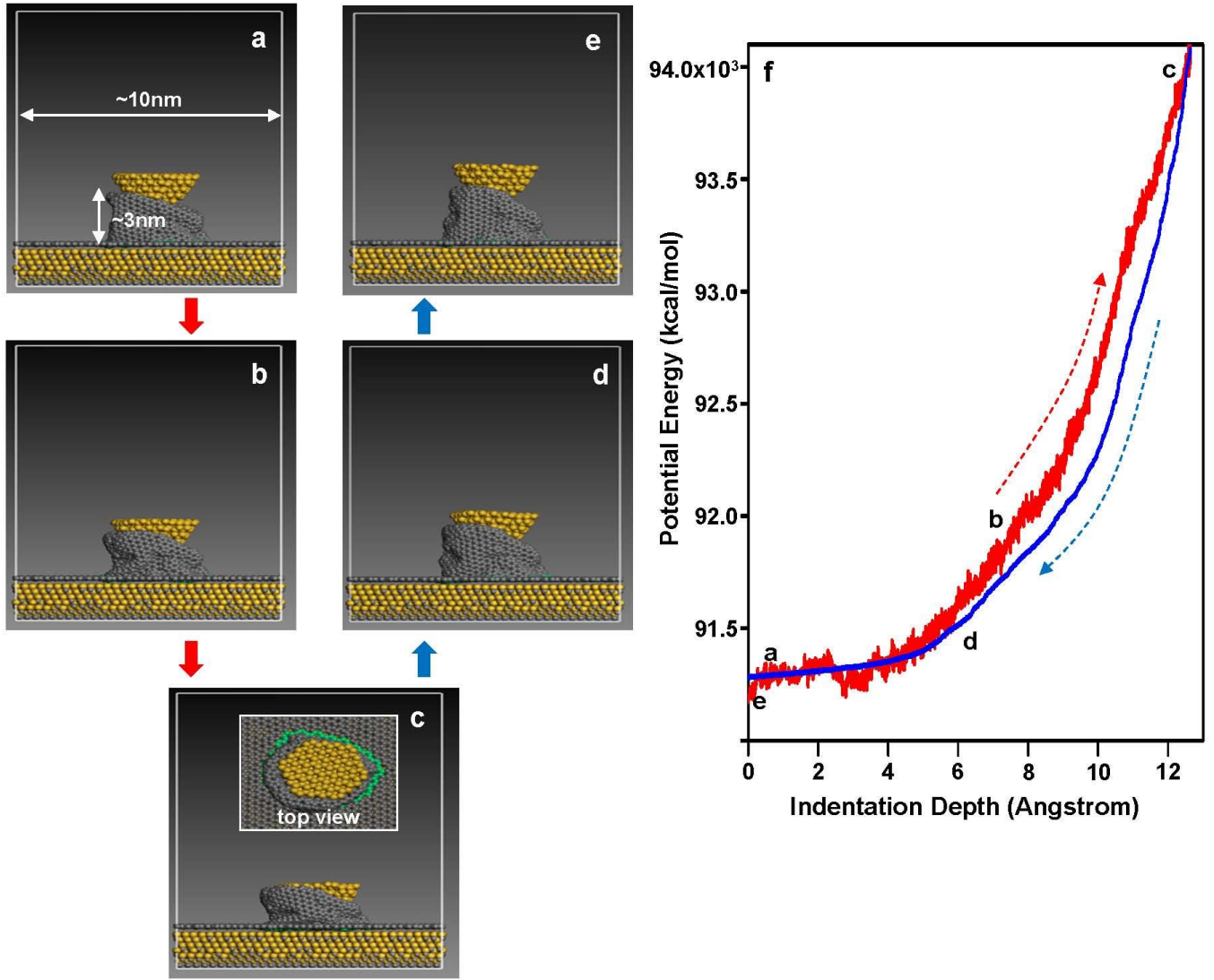

**Figure 6 | Variation of potential energy in a 3D freestanding graphene model with external loading.** (**a-e**) Snapshots during the loading-unloading MD simulation, where the yellow cone-shaped Si tip moves up and down, are shown on the left. The gray atoms are carbon. (**f**) Points corresponding to the snapshots are indicated on the potential energy-displacement curve on the right. Note that the potential energy is that of only the freestanding graphene. The total indentation depth is approximately 1.3 nm. The red dotted arrow represents downward loading and the blue dotted arrow represents upward unloading.

*Supplementary Information*

# Designed Three-Dimensional Freestanding Single-Crystal Carbon Architectures

Ji-Hoon Park[1]**, Dae-Hyun Cho[2]**, Youngkwon Moon [1], Ha-Chul Shin[1], Sung-Joon Ahn[1], Sang Kyu Kwak[3], Hyeon-Jin Shin[4], Changgu Lee[2,5]*, and Joung Real Ahn[1,5]*

[1]Department of Physics, Sungkyunkwan University, Suwon, 440-746, Republic of Korea

[2]Department of Mechanical Engineering, Sungkyunkwan University, Suwon, 440-746, Republic of Korea

[3]School of Energy and Chemical Engineering, Ulsan National Institute of Science and Technology (UNIST), Ulsan 689-798, Republic of Korea

[4]Graphene Center, Samsung Advanced Institute of Technology, Yongin, 446-712, Republic of Korea

[5]SKKU Advanced Institute of Nanotechnology, Sungkyunkwan University, Suwon, 440-746, Republic of Korea

**Received date (automatically inserted)**

\* To whom correspondence should be addressed: E-mail: peterlee@skku.edu(CL), jrahn@skku.edu(JRA),

\*\*These authors contributed equally to this work.

**S1. Optical images of 3D SiC architectures after heating**

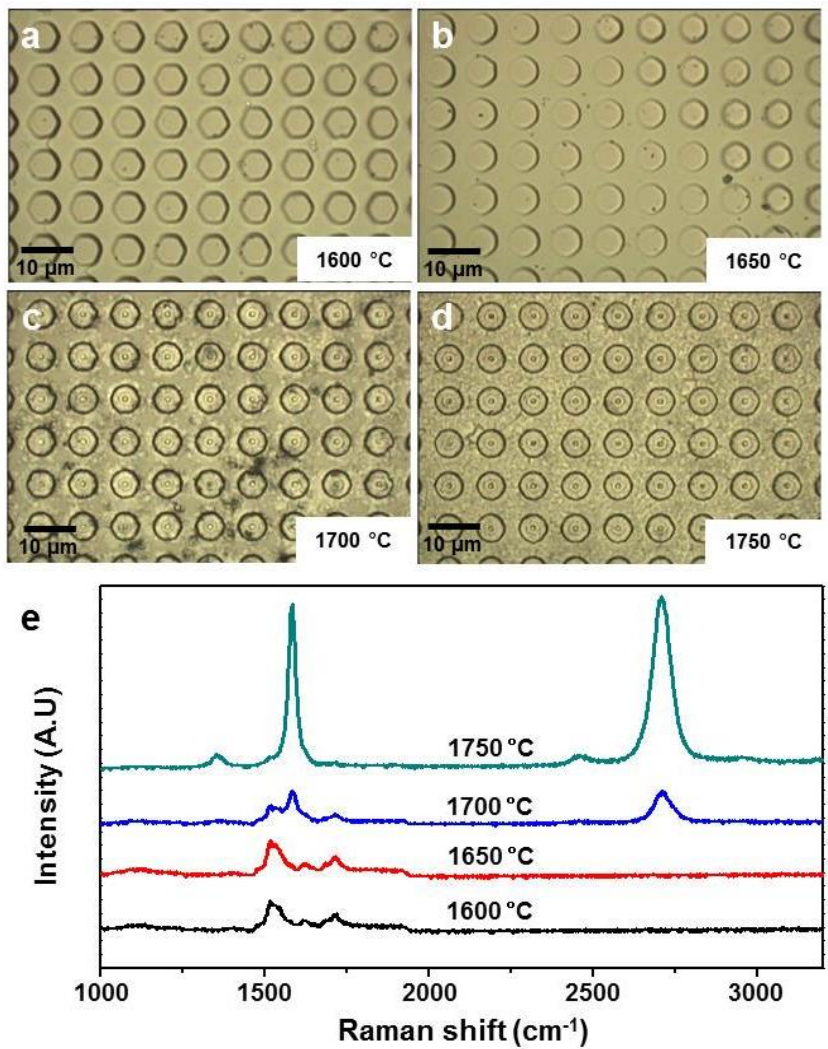

**Figure S1**. **(a-d)** Optical images acquired after heating the inverted bowl-shaped SiC architectures at various temperatures: (**a**) 1600°C, (**b**) 1650°C, (**c**) 1700°C, and (**d**) 1750°C. (**e**) Raman spectra obtained after heating the SiC architectures.

**S2. The effects of oxygen plasma on a SiC template**

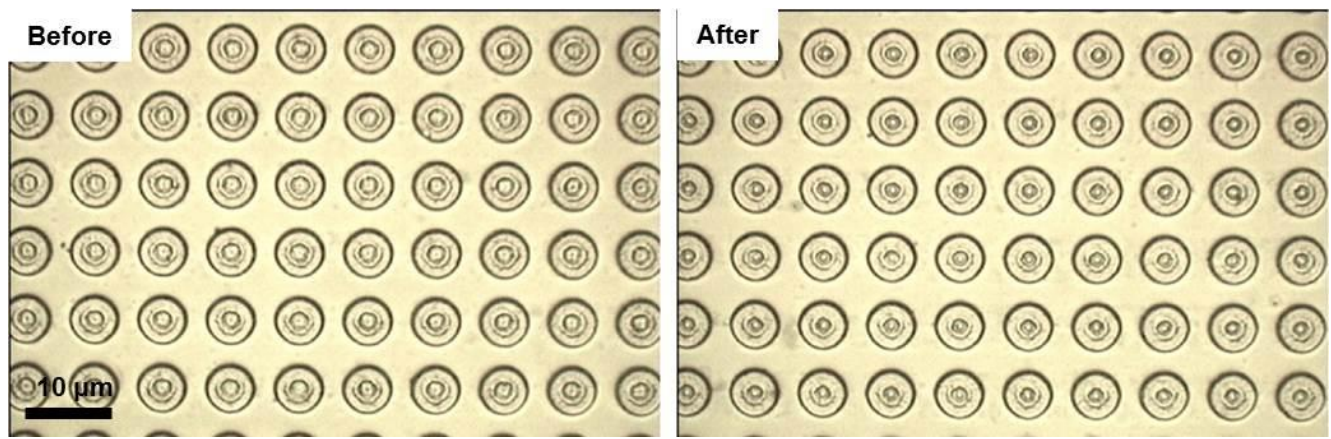

**Figure S2**. Optical images of a SiC template before and after oxygen plasma treatment.

**S3. Raman spectra taken during the growth of 3D freestanding graphene architecture**

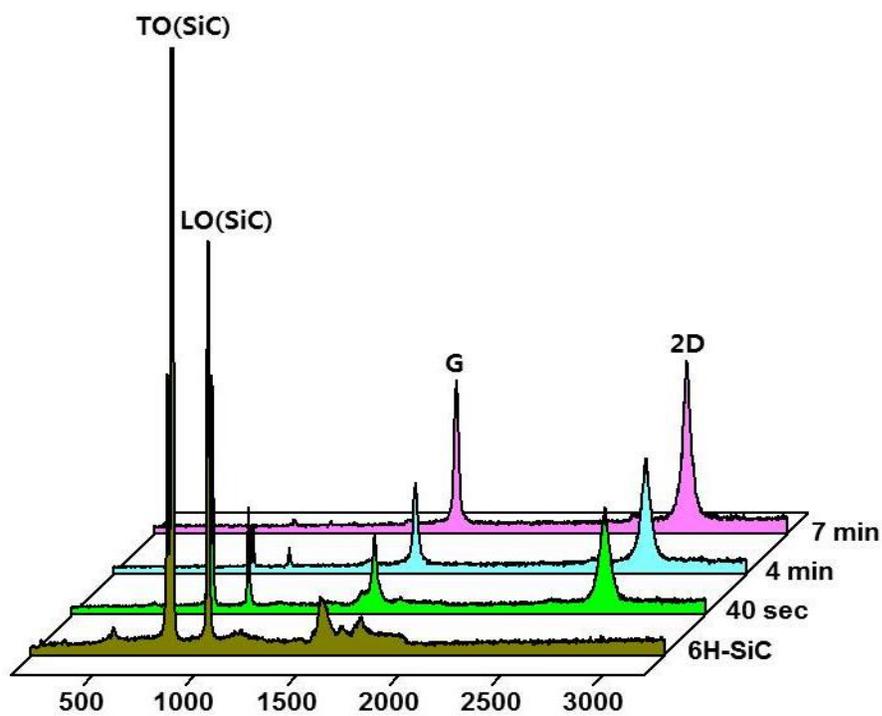

**Figure S3**. The Raman spectra were acquired from a designed SiC template (khaki) before heating and from graphene-covered SiC structures after heating for 40 seconds (green), 4 minutes (blue), and 7 minutes (pink) at 1750°C under an argon atmosphere.

## S4. Mechanical properties of graphene-covered SiC structures

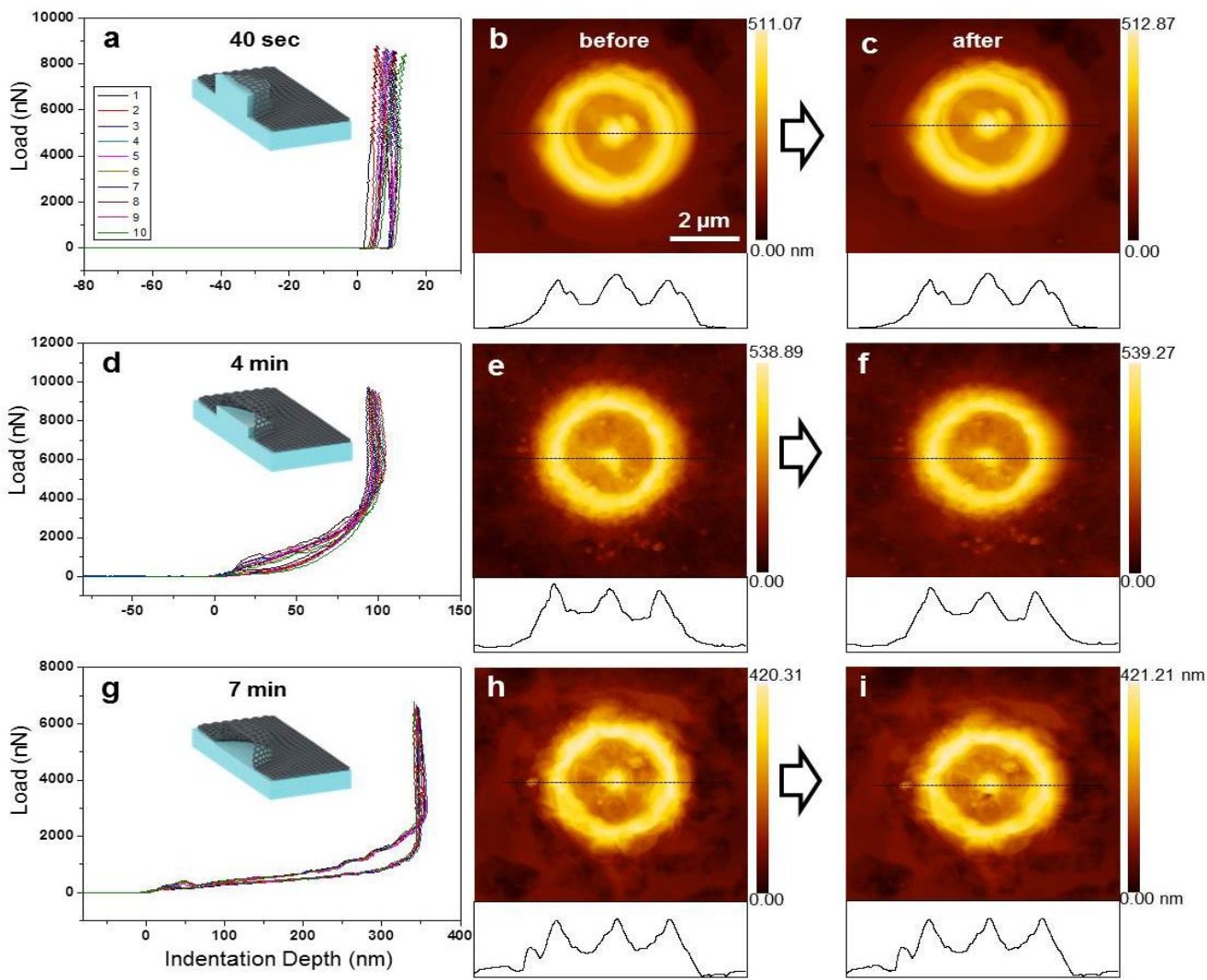

**Figure S4**. (**a**) Load-displacement curves for the first 10 cycles of a graphene-covered SiC structure heated for 40 seconds at 1750°C under an argon atmosphere. (**b-c**) The AFM topographies and line profiles before (**b**) and after (**c**) the indentations in **a**. (**d-i**) The mechanical properties of the graphene-covered SiC structures heated for 4 (**d-f**) and 7 (**g-i**) minutes under the same conditions.

**S5. Loading speed dependence**

For carbon nanotubes (CNTs), it has been reported that both viscoelasticity and shell buckling strongly influence their elastic hysteresis.[1-5] According to measurements of the mechanical responses of multiwall CNTs, the dissipation of energy is caused by the coupling between neighboring walls, including the van der Waals interactions between the π electrons of the inner walls, and the friction or shear resistance between the inner and outer nanotubes.[1,2,5] Furthermore, other entangled materials, such as eukaryotic cells with protein bonds, exhibit hysteretic elasticity responses in loading-unloading cycles.[6] For graphene, a slip between graphene layers is predicted to happen under biaxial tensile stress.[7] Lu *et al*. calculated the critical strains for the sudden onset of buckling for monolayer graphene ribbons under uniaxial compression.[8] Among the possible mechanisms of the hysteresis exhibited by the 3D freestanding graphene architecture, viscoelasticity can be ruled out because the responses to the indentations are independent of loading speeds in the range 6.7 to 4000 nm/s. Therefore, the buckling phenomenon and/or the decoupling of graphene layers are possible origins of the hysteresis loop. When the AFM tip indents and exerts normal stress on the center of the graphene architecture, the strut part undergoes bending and compression. In contrast to the strut, tensile and bending stresses are exerted on the flat part. The stresses that are exerted on the architecture can result in different elongations in the outer and inner layers of graphene. When the elongation difference exceeds 0.14 nm, which is equal to the carbon-carbon bond length, a sudden slip between graphene layers can begin in some parts of the graphene architecture,[7] which results in fully recoverable buckling.[5] From the hysteresis loop of the 3D freestanding graphene architecture, we suggest that the critical elongation difference occurs at a load of approximately 500 nN, which corresponds to the buckling load.

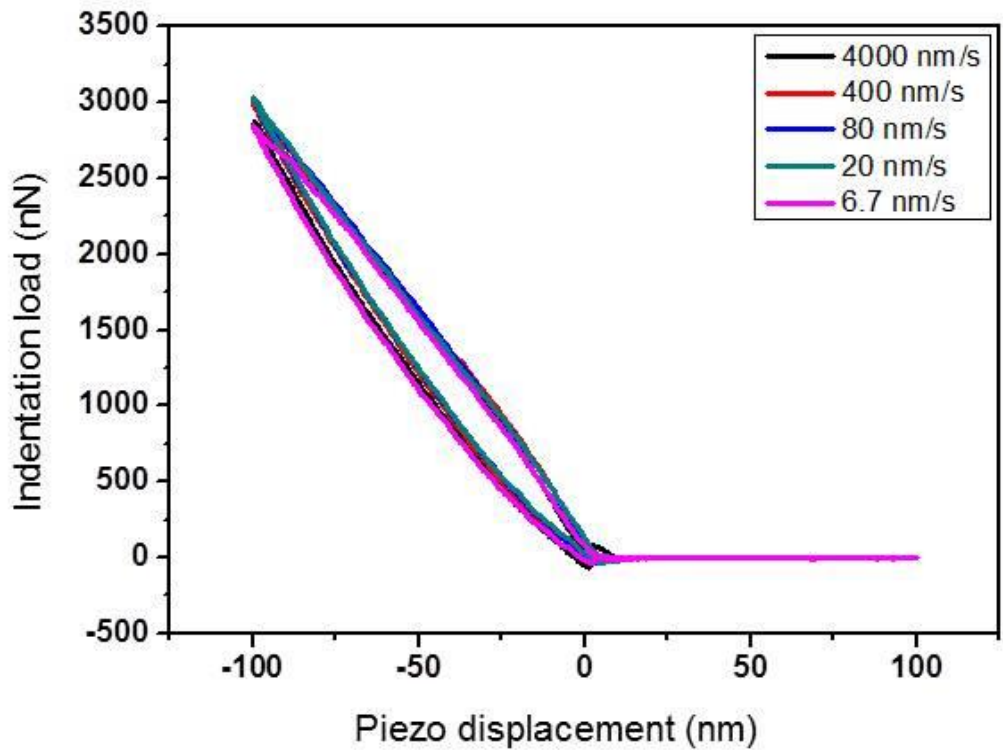

**Figure S5**. Load-displacement curves for the inverted bowl-shaped 3D freestanding graphene architecture. Loading speeds were varied from 6.7 to 4000 nm/s.

**S6. 3D freestanding graphene architecture transferred onto a SiO₂/Si wafer**

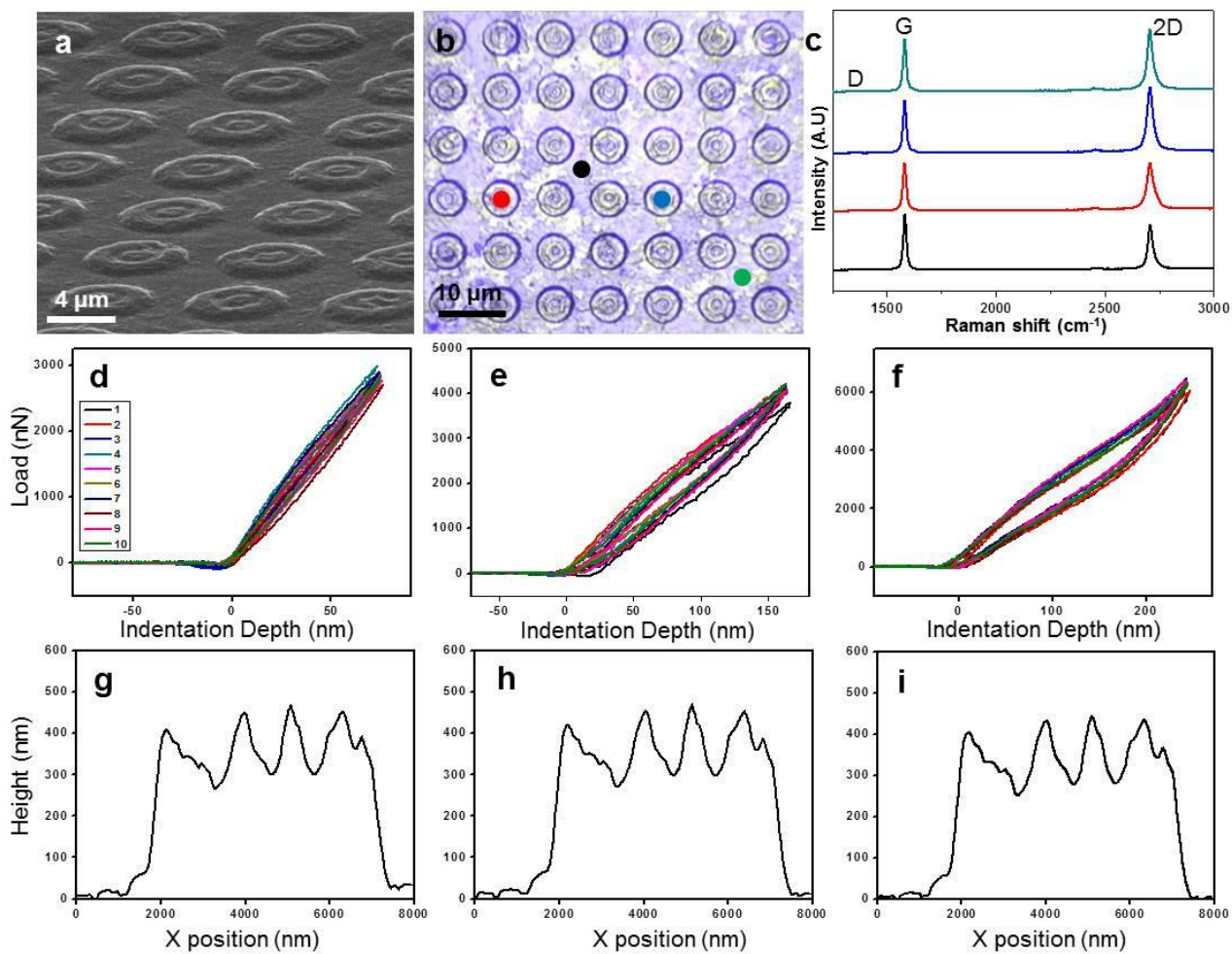

**Figure S6**. (**a**) SEM image of 3D freestanding graphene architecture transferred onto a SiO$_2$/Si wafer. (**b**) Optical image of the graphene architecture. (**c**) Raman spectra of the graphene architecture obtained at the colored dots in **b**. (**d-f**) Load-displacement curves during the first 10 cycles at indentation depths of 70 nm (**d**), 170 nm (**e**), and 230 nm (**f**). (**g-i**) Line profiles acquired after indentations. The profile was taken across the center of the inverted bowl structure.

## S7. Molecular dynamics simulations

In the MD simulations in Figure 6, the molecular weight of Si has been increased 1000 times to mimic the loading of the AFM tip so that the cone does not deviate from a vertical trajectory during the simulation. The loading and unloading speeds of the cone are estimated to be approximately 81 and 0.98 Å/ps, respectively, during microcanonical (i.e. constant $NVE$) MD runs, where $N$ is the number of atoms, $V$ is the volume, and $E$ is the energy, with a time step of 1 fs. For the MD simulations, the COMPASS force field of the Materials Studio (version 6.1) program was used. [http://accelrys.com/products/materials-studio/]